# FIPS Compliant Quantum Secure Communication Using Quantum Permutation Pad

Alex He
Quantropi Inc.
Ottawa, Canada
alex.he@quantropi.com

Dafu Lou
Quantropi Inc.
Ottawa, Canada
dafu.lou@quantropi.com

Eric She
DLS Technology Corporation
Ottawa, Canada
eshe@dlstech.com

Shangjie Guo
FinQ Tech Inc.
College Park, Maryland
sguo@finq.tech

Hareesh Watson
DLS Technology Corporation
Ottawa, Canada
hwatson@dlstech.com

Sibyl Weng
DLS Technology Corporation
Ottawa, Canada
sweng@dlstech.com

Maria Perepechaenko
Quantropi Inc.
Ottawa, Canada
maria.perepechaenko@quantropi.com

Randy Kuang
Quantropi Inc.
Ottawa, Canada
randy.kuang@quantropi.com
ORCID: 000-0002-5567-2192

*Abstract*—Quantum computing has entered a fast development track since Shor's algorithm was proposed in 1994. Multi-cloud services of quantum computing farms are currently available. One of which, IBM quantum computing, presented a road map showing their Kookaburra system with over 4158 qubits that will be available in 2025. For the standardization of Post-Quantum Cryptography or PQC, the National Institute of Standards and Technology or NIST recently announced the first candidates for standardization with one algorithm for key encapsulation mechanism (KEM), Kyber, and three algorithms for digital signatures. NIST has also issued a new call for quantum-safe digital signature algorithms due June 1, 2023. This timeline shows that FIPS-certified quantum-safe TLS protocol would take a predictably long time. However, 'steal now, crack later' tactic requires protecting data against future quantum threat actors today. NIST recommended the use of a hybrid mode of TLS 1.3 with its extensions to support PQC. The hybrid mode works for certain cases but FIPS certification for the hybridized cryptomodule might still be required. This paper proposes to take a nested mode to enable TLS 1.3 protocol with quantum-safe data, which can be made available today and is FIPS compliant. We discussed the performance impacts of the handshaking phase of the nested TLS 1.3 with PQC and the symmetric encryption phase. The major impact on performance using the nested mode is in the data symmetric encryption with AES. To overcome this performance reduction, we suggest using quantum encryption with a quantum permutation pad for data encryption with a minor performance reduction of less than 10%.

I. INTRODUCTION

Peter Shor proposed his celebrated quantum algorithm in 1994 [1], which solves the NP-hard problem of prime integer factorization in polynomial time. At its beginning, quantum computing, especially universal gate-based quantum computing, experienced a slow development phase for about two decades. In 2019, Arute et al. from Google claimed Quantum Supremacy with their 53-qubits Sycamore processor [2]. This marked the start of the global quantum computing race. Since IBM released their 5-qubit quantum computer for public access with Qiskit tool in 2017, IBM recently announced their 433-qubit quantum computer and plan to double its qubits every year for 2023 to reach over 4,000-qubits by 2025, outlined in their development roadmap [3].

The fundamental shift from classical computing to quantum computing is the shift in computing algebra from the classical Boolean algebra to linear algebra, used in quantum computing. That is, from classical logic gates, implemented in CPU, to quantum logic gates to be implemented in QPU. That means, quantum computing indicates a quadratic speedup of classical computers or $(2^n)^2$, where $n$ denotes the number of qubits of a QPU. This exponential computing power would break today's RSA-2048 in 10 seconds if a fault tolerant quantum computer reaches 4099 qubits, while a classical computer would take 300 trillion years. Mosca predicted that there is a "1/2 chance of breaking RSA-2048 by 2031" [4]. The year from quantum computing to break classical public key RSA has been called the Year to Quantum threat or Y2Q.

Symmetric cryptography such as the well-known Advanced Encryption Standard or AES also suffered quadratic speedup of the best quantum attack on the key space using Grover's algorithm proposed by Grover in 1996 [5]. That requires the key length to be doubled in comparison with the equivalent classical security level. For example, the classical AES-128 would be replaced by AES-256 for quantum security.

It has been well-understood that the upcoming quantum computing systems will destroy the foundation of classical public key infrastructure or PKI for both key establishment such as RSA, Diffie-Hellman or elliptical curve Diffie-Hellman and digital signature such as Digital Signature Algorithm or DSA. National Institute of Standards and Technology or NIST has announced its standardization process of Post-Quantum Cryptography or PQC in November of 2017. In July 2022 [6], NIST announced the lattice-based Kyber [7] to be its first standardized key encapsulation mechanism or KEM. And lattice-based Dillithium [8], Falcon [9], as well as hash-based SPHINCS+ [10] to be first standardized digital signature schemes. Other KEM candidates BIKE [11], Classic McEliece [12], HQC [13], and SIKE [14] moved to the 4th round to be considered further. NIST has also announced its reopening submissions for digital signature standardization due June 2023 [15].

In 2022 some PQC algorithms such as Rainbow (a digital signature scheme based on Multivariate Public Key Cryptography or MPKC) [16] and Supersingular Isogeny Diffie-Hellman protocol or SIDH [17, 18] proved to be vulnerable to classical attacks. Another very interesting cryptoanalysis was reported in the late 2022 by Wenger et al. [19]. The team used transformers (a deep learning model) to develop an attack on certain lattice-based schemes. They noticed that the basic equation system for Learning With Error or LWE can be expressed as linear regression used in machine learning. If trained transformers, especially combined with quantum computational advantage, can solve the Short Vector Problem, then PQC algorithms face an enormous challenge. Recall that most of the PQC standard schemes are lattice-based. This issue is especially concerning since estimation on Y2Q has not taken the power of quantum machine learning into account.

Some recent novel PQC algorithms were proposed by Kuang, Perepechaenko, and Barbeau for KEM [20, 21] and digital signature [22, 23, 24, 25], based on NP-complete Modular Diophantine Equation Problem or MDEP. These novel schemes share a foundation which we call Multivariate Polynomial Public Key or MPPK. MPPK is built on two vector spaces: a linear multivariate vector space $\{x_1, \ldots, x_m\}$ containing noise variables used for obscurity and polynomial vector space $\{1, x^1, \ldots, x^{n+\lambda}\}$ containing secret message variable. MPPK offers small parameter sizes at the level of hundreds of bytes for key, ciphertext, and signature and outperforms NIST finalists in key generation, encryption, decryption, as well as signing and signature verification. They could be considered as generic PQC algorithms for a wide range of devices and systems.

NIST recommended using a hybrid scheme for quantum resistant TLS 1.3 [26]. By leveraging the keyShare extension of TLS 1.3, one can combine a NIST-approved classical key establishment algorithm such as ECCDH in TLS 1.3 with one or more PQC KEM algorithms and a NIST-approved digital signature algorithm such as DSA with a list of PQC digital signature algorithms for a chain signatures. This hybrid scheme offers crypto agility as the ongoing process of PQC standardization and cryptanalysis. However, this hybrid scheme brings two potential limitations: 1) the hybrid TLS 1.3 may still require FIPS certification although its core cryptomodule is certified in classical TLS 1.3. In general, the FIPS certification comes at a cost for both time and money, although it would be possible to receive the certificate; 2) The certified hybrid TLS 1.3 must be integrated with an application for a quantum resistant service. But in some cases, this integration may be difficult for some applications running inside web browsers.

This paper proposes a new hybrid scheme by nesting PQC inside classical TLS 1.3, creating nested TLS 1.3, to overcome the limitations in the above hybrid scheme. The nested TLS 1.3 does not require a new FIPS certification because it does not change the existing certified TLS. Moreover, it supports TLS 1.3 as well as any certified TLS such as TLS 1.2 or even earlier. One drawback of the nested scheme is that it may reduce the performance of the encryption and decryption operations because the transmitted data would be encrypted twice if AES is used for both encryptions. To minimize the performance impact, we propose to use quantum encryption with a Quantum Permutation Pad algorithm or QPP [27, 28]. The QPP will be used to encrypt the raw data first, producing a quantum-encrypted message used as a TLS 1.3 message to be encrypted with AES.

## II. FIPS COMPLIANT QUANTUM ENCRYPTION WITH QPP

In this section, we first introduce the concept of a nested TLS 1.3 protocol, that uses quantum-safe cryptosystems. The nested TLS 1.3 allows for a smooth transition from classical era to quantum era while maintaining FIPS compliance. We then discuss the nested TLS 1.3 handshake process. Next, we consider symmetric data encryption with QPP for consideration of both performance and security.

### A. Nested TLS 1.3 with PQC in TLS Handshaking Proccess

The proposed nested TLS 1.3 is illustrated in Fig. 1. The figure illustrates an Open Systems' Interconnection model (OSI-model) consisting of 7 layers: physical, data link, internet, transport, session, presentation, and application. The functionality of each layer is illustrated in [29]. On the right-hand side, we mark the corresponding layers in TCP/IP model. The existing TLS cryptomodule is in the transport layer, while the nested TLS is in the application layer. White arrows denote a FIPS certified TLS 1.3 for clientHello request from the client and serverHello response from the server to establish a shared session for session data encryption and decryption. The NIST-Approved key agreement protocol, ECCDH, is used for forward secrecy in the existing TLS. The RSA algorithm is excluded from the key agreement protocol. RSA, DSA, and ECDSA are paired with hash functions for digital signature in the existing TLS. The current conventional FIPS certified TLS cryptomodule will be vulnerable to quantum computing attacks once the fault tolerable quantum computers are available. However, the "steal now crack later" tactics are already in use, meaning all encrypted data today is at risk. Immediate action is imperative to protect data against future quantum threat actors. If sensitive information is required to remain secret for over 10 years, then it would not be wise to wait for the FIPS certified TLS 1.3 with quantum resistance.

The proposed nested TLS 1.3 with PQC cryptographic modules is independent from the FIPS certified TLS cryptomodule in a sense that packets from the nested TLS 1.3 become data packets for the FIPS certified TLS. Since the outer classical TLS cryptomodule is not altered, the nested TLS 1.3 does not violate the FIPS certification. This solution can be considered as a promising FIPS compliant TLS 1.3 for quantum security. The nested TLS 1.3 can be used to turn "steal now, crack later" into "steal now, safe forever".

Nested TLS 1.3 is based on the Open Quantum Safe or OQS OpenSSL to support PQC KEM algorithms such as the NIST finalists Kyber and Saber, as well as MPPK [20] and Homomorphic Polynomial Public Key further evolved from MPPK [30]. For the digital signatures, nested TLS 1.3 supports PQC digital signature algorithms such as NIST finalists Falcon,

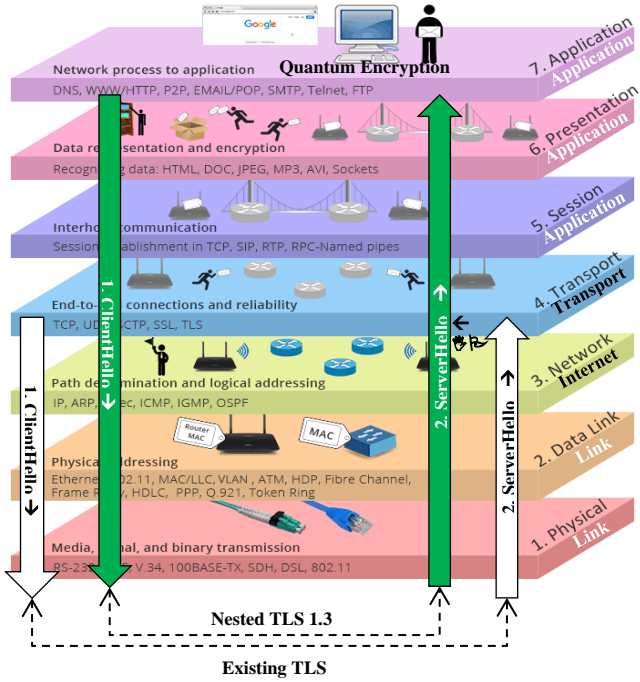

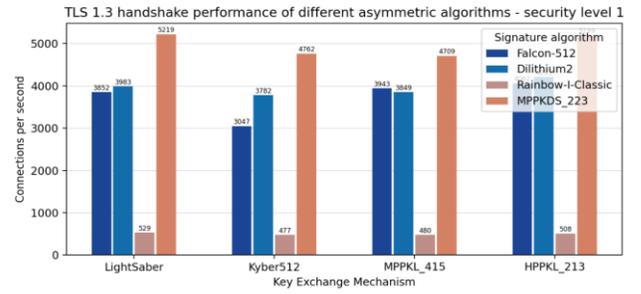

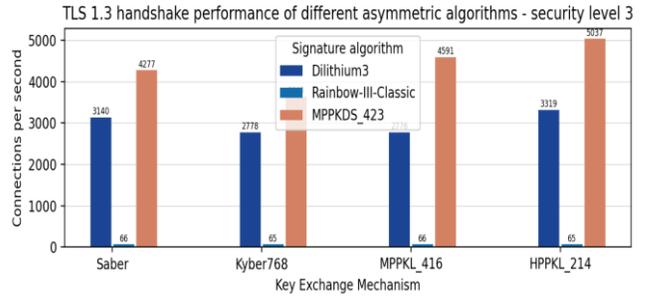

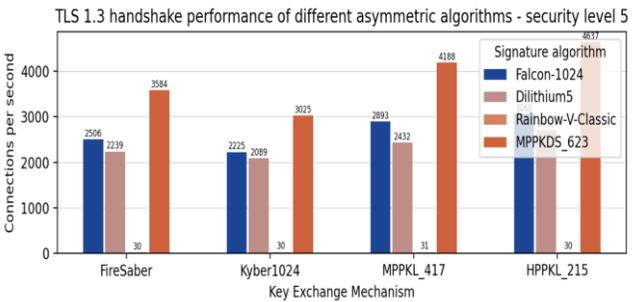

Fig. 1. TLS 1.3 handshake performances (connections/second) are illustrated in terms of PQC KEM algorithms paired with PQC digital signature algorithms. The x-axis is marked with KEM schemes and y-axis represents handshaking connections per second. Digital signature schemes are listed in the legends. NIST security level I, III, and V are associated with the top, middle, and bottom graphs respectively.

Dilithium, Rainbow, as well as MPPK/DS to be submitted to NIST for standardization in 2023 [22] or its new variants.

### B. Performance of a Nested TLS 1.3 Handshake

We tested the performance of a nested TLS1.3 handshake in a local machine on a 16-core Intel®Core™ i7-10700 CPU at 2.90 GHz system for all the measured primitives. Fig. 2 illustrates the performance of TLS 1.3 handShake for each pair of KEM and digital signature schemes in terms of NIST security levels. In general, Rainbow digital signature demonstrates the worst performances for TLS handshake with about 500 connections/second at all NIST security level I, 65 connections/second at all NIST security level III, and 30 connections/second at all NIST security level V.

By pairing MPPK/DS with Saber, Kyber, MPPK KEM, and HPPK, the MPPK/DS scheme outperforms digital signature schemes Falcon and Dilithium over 30% at security level I, over 35% at security level III, and 40% at security level V, respectively. On the other hand, pairing MPPK KEM and HPPK with NIST digital signature finalists Falcon and Dilithium, as well as MPPK/DS, the pairs of MPPK KEM with MPPK/DS and HPPK with MPPK/DS outperform NIST finalists Falcon and Dilithium. HPPK pairing with MPPK/DS demonstrates a slightly better performance than MPPK KEM pairing with MPPK/DS, about 10% for all three security levels.

Fig. 2 also demonstrates that the average TLS 1.3 handShake can be completed at sub-million seconds. For example, MPPK/DS paired with MPPK KEM and HPPK would establish about 5000 TLS 1.3 connections per second which gives 0.2 ms/connection. In an actual cloud environment, the performance would be reduced due to the network latency. A typical network latency is at 10 ms level, so a sub-million second processing time contributes no impact on a practical TLS 1.3 handShake. That means, a nested TLS 1.3 inside the existing TLS cryptomodule would not impact the overall performance if we consider the fact that there is only one handshaking per session.

Fig. 2. Illustration of nested TLS with PQC and quantum encryption in OSI and TCP/IP models. The colorful OSI model is taken from literature [29]. The TCP/IP model is indicated on the right-hand side. The white arrows refer to the existing certified TLS 1.x and green arrows denote the nested TLS 1.3 with quantum resistant cryptographic modules.

### C. Nested TLS 1.3 with PQC in Symmetric Encryption

After the handshake process of a nested TLS is complete, communication peers establish a shared session key for symmetric encryption during the session. Undoubtedly, the NIST-Approved AES-256 can be used for data encryption in the nested TLS, and the produced ciphertext would be encrypted again by the outer FIPS certified AES-256 with its own session key. In this case, the session keys may be potentially obtained by attackers using "Steal now, crack later" strategy, and later decrypted using quantum attacking mechanisms. However, if the nested TLS uses quantum-safe algorithms for encryption, then vulnerability of the outer session does not influence the security of transmitted data to a great extent. That is, if the nested TLS 1.3 with PQC establishes a secure session key for session data encryption, then the attacker with quantum resources will not be able to decrypt the data encrypted in the nested TLS layer even if they were able to decrypt the data encrypted in the outer classical layer. This nested TLS 1.3 with PQC stops the "steal now, crack later" and offers the "steal now, safe forever".

If AES-256 is used for encryption in conventional outer and nested layer, that would cause the overall performance to drop by 50%. However, there is no requirement to use AES-256 for the nested layer encryption. Under the consideration of the FIPS compliance, the inner data encryption does not need to be the NIST-Approved and FIPS certified, we can choose quantum encryption with Quantum Permutation Pad or QPP [28], implemented classically with permutation matrices. Unlike AES-256 encryption with 14 rounds, QPP encryption follows the same way as quantum gate operations or matrix vector multiplications. QPP encryption is bijective transformation so typical pre-randomization and dispatch techniques would be applied before the gate operations to avoid statistic patterns appearing in the ciphertext. Quantum encryption with QPP demonstrates excellent performance in both encryption and decryption, being over 10x faster than AES-256 [31, 32, 33, 34] Using QPP algorithm for encryption in the inner layer does not impact the performance as greatly as AES-256. The drop in performance with QPP is less than 10%. At this cost we can offer quantum secure TLS with "steal now, safe forever". In addition, quantum encryption with QPP has been implemented into IBM quantum computers and compiled into 2-qubit and 3-qubit quantum circuits [35, 36, 37, 38].

For the detailed discission of quantum encryption with QPP, please refer to [27-34]. Here we briefly summarize classical or quantum implementation of QPP as follows shown in Fig. 3:

1. Choose the number of n-bits or n-qubits used to generate the permutation gates, for example, $n = 4, 8$.

2. Choose the number of gates to be used, M. M=64 for $n = 8$ in the digital QKD implementation [39]. It can be reduced to 8 permutation gates with $n = 4$.

3. Session key is first expanded for M permutation gates and then mapped to a set of permutation gates through the Init module where the Fisher-Yates shuffling algorithm.

4. The session key is also used to seed a cryptographic pseudo random number generator or PRNG

5. The pseudo random number generated by PRNG is used to pre-randomize the plaintext m with XOR and then dispatch the randomized data to the indexed permutation gate for encryption and then the ciphertext c is output accordingly.

6. At the receiving side, the process is symmetric to the encryption side, but the permutation gate must be reversed or transposed. The pseudo random number is first used to dispatch the ciphertext to the indexed permutation for decryption and after then derandomize with the pseudo random number, the original plaintext m is obtained.

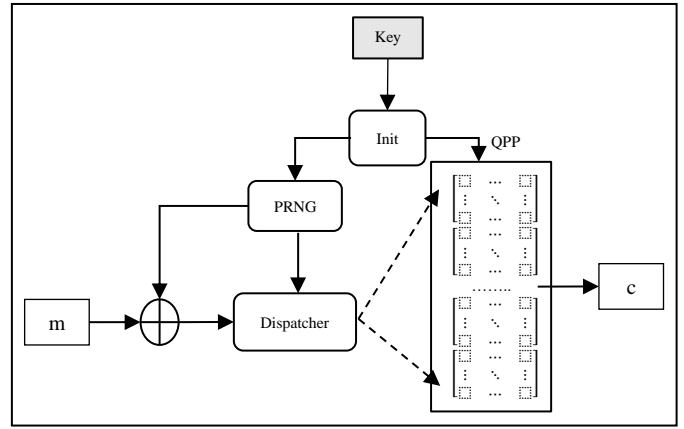

Fig. 3. Quantum encryption with QPP is illustrated with the session key, plaintext data, and ciphertext together with related modules.

From Fig. 3, we can see that quantum encryption with QPP only consists of three steps to complete its encryption: randomization to remove any statistic bias, dispatching to the indexed permutation gate/matrix, and then gate operation applied to the randomized plaintext. Overall, the entire encryption process may take at most the process time of a single round in AES encryption. That is why QPP would be 10x faster than AES. Therefore, the nested TLS 1.3 with PQC only has a minor impact on the communication performance, less than 10%.

If the attackers apply the "steal now, crack later" strategy and wait for the quantum computers to break the public key for the session key, then they can decrypt the outer AES encryption and obtain the quantum encrypted ciphertext. The quantum encrypted ciphertext is then required to be cracked. However, QPP as well as PQC algorithms and any other quantum-safe algorithms are designed to withstand classical and quantum attacks.

We have performed randomness analysis on ciphertext produced by QPP. Table I illustrates the randomness analysis of very biased English character files and ciphertext encrypted with QPP using ENT test tool. ENT randomness test tool is very sensitive to bit and byte level bias, especially detected in the Chi Square values. ENT outputs six reports on their entropy per 8 bits, Chi Square value, p-value, arithmetic mean, Monte Carlo $\pi$, and serial correlation. Table I shows that the total biased English plaintext is encrypted with QPP into ciphertexts which demonstrate excellent randomness, especially the Chi Square value 233.2 with a p-value 0.83. The acceptable p-value a good randomness is from 0.01 to 9.99. The ciphertext also demonstrated excellent value for arithmetic mean 127.49, Monte Carlo $\pi = 3.14198164$, and finally the serial correlation at $9.3 x 10^{-5}$.

Indeed, the nested TLS 1.3 with PQC offers FIPS compliant solution with a quantum encryption component. We this this solution as a good strategy for transition from classical security to quantum security without waiting for NIST standardization and FIPS certification to complete necessary processes. Crypto agility is essential nowadays, since PQC and other algorithms are novel and might have undiscovered attacks, especially as quantum computing matures. So, whenever a specific algorithm is found to be vulnerable, the algorithm can be easily removed from the nested cryptomodule and replaced with a new one in a convenient and quick manner. When all PQC KEMs and digital signature algorithms are standardized and the FIPS certification is required, then the whole TLS 1.3 cryptomodule would be used for FIPS certification. Once it is FIPS certified, the nested TLS 1.3 automatically becomes certified quantum resistant TLS 1.3 cryptomodule to replace the classical TLS 1.3. With this, we feel confident to turn the "steal now, crack later" into "Steal now, safe forever".

TABLE I. ENT Testing Is Tabulated for Statistically Biased Plaintext Inputs and Ciphertext Encrypted with QPP, together with Their Ideal Values

| ENT | Plaintexts | Ciphertext | Ideal Value |
|---|---|---|---|
| Entropy (bits) | 4.224280 | 7.999998 | 8.000000 |
| Chi Square | 1821992676 | 233.20 | 256 |
| p-Value | 0.0001 | 0.83 | 0.5 |
| Arithmetic Mean | 97.9686 | 127.4953 | 127 |
| Monte Carlo $\pi$ | 4.000000000 | 3.14198164 | 3.14159265 |
| Serial Correlation | -0.138722 | - 0.000093 | 0 |

### III. CONCLUSION

In this work, we propose a FIPS compliant TLS 1.3 solution with a nested quantum-secure TLS component. This solution makes seamless transition and mitigation from classical security to quantum security, that does not require any wait time for standardization and certification. Given that the standardization and FIPS certification process might take 10 years, adversaries can use this time to take advantage of the "steal now, crack later" strategy. The proposal of the nested TLS 1.3 with quantum-safe component could turn the "steal now, crack later" into "steal now, safe forever", while preserving FIPS certified outer TLS layer. Therefore, this proposed work is critical for protecting sensitive data today with long shelf life against future quantum threats, which may impact sectors including public health, insurance, genetics, retirement. Any symmetric algorithm can be used in the nested TLS layer. However, to overcome performance impact in symmetric encryption, we suggest using quantum encryption with QPP to further enhance the data security even with the successful crack the outer public key with quantum computer, the inner TLS 1.3 is still secure. In the future, we plan to build the nested TLS 1.3 with PQC and test its real performance in a cloud environment in comparison with normal TLS 1.3 with PQC.

### ACKNOWLEDGMENT

We acknowledge that Ryan Toth provided the OQS OpenSSL diagram shown in Fig. 2. The overall performance of TLS 1.3 hand shaking with MPPK/DS and MPPK KEM, as well as HPPK would be published separately.

We acknowledge that the image in Fig. 1 was originally taken from [29]. The image is licensed under the Creative Commons Attribution 4.0 International [40]. We have made modifications to the original image available in its original form in [29].